Emergent phenomena in living systems: a statistical mechanical perspective

Indrani Bose
Department of Physics
Bose Institute
Kolkata, India
E-mail: indrani@jcbose.ac.in
ibose1951@gmail.com

Abstract. A natural phenomenon occurring in a living system is an outcome of the dynamics of the specific biological network underlying the phenomenon. The collective dynamics have both deterministic and stochastic components. The stochastic nature of the key processes like gene expression and cell differentiation give rise to fluctuations (noise) in the levels of the biomolecules and this combined with nonlinear interactions give rise to a number of emergent phenomena. In this review, we describe and discuss some of these phenomena which have the character of phase transitions in physical systems. We specifically focus on noise-induced transitions in a stochastic model of gene expression and in a population genetics model which have no analogs when the dynamics are solely deterministic in nature. Some of these transitions exhibit critical-point phenomena belonging to the mean-field Ising universality class of equilibrium phase transitions. A number of other examples, ranging from biofilms to homeostasis in adult tissues, are also discussed which exhibit behavior similar to critical phenomena in equilibrium and nonequilbrium phase transitions. The examples illustrate how the subject of statistical mechanics provides a bridge between theoretical models and experimental observations.

1. Introduction

The living cell is an example of an interacting many body system in which the genes are not independent entities but constitute a complex network of interactions. In a genetic network the nodes represent the individual genes. A pair of nodes is connected by a link if the protein product of one gene regulates (activates/represses) the expression of the other gene. Besides the genetic network, one can define other types of biological networks like the protein-protein interaction and metabolic networks (Barabási and Oltvai 2004). Depending on the specific cellular event, one needs to consider only a sub-network of the full network. The living cell may be represented as a dynamical system in which a myriad of biochemical events unfold as a function of time. The networks thus have a structure as well as dynamics. In a deterministic description of the dynamics, one solves a set of differential rate equations to determine how the key variables defining the dynamics evolve as a function of time. A fully deterministic description, however, ignores the inherent stochastic nature of cellular events like cell division, cell differentiation and gene expression (GE). We choose the example of GE, a central activity in the living cell, to discuss how stochasticity influences the fundamental cellular process (Kærn et al. 2005, Raj and van Oudenaarden 2008). The different biochemical events involved in GE have a natural description in terms of chemical reactions. A chemical reaction is inherently probabilistic in nature and occurs when the reactants come together through random collisions with the molecules in the surrounding medium, making the time of occurrence of the reaction uncertain. This is more so if the number of reactant molecules is small. The initiation of transcription, the first step in GE, is preceded by a binding reaction involving a RNA polymerase (RNAP) molecule and a DNA. Transcription is thus initiated at different times in the individual cells of a cell population. The stochasticity arising from single cell chemical kinetics is intrinsic in nature. The extrinsic origin of stochasticity lies in the cell-to-cell variations in the numbers of biomolecules involved in GE e.g., the transcription factors (TFs) and the RNAP molecules. When GE dynamics is solely deterministic, a single protein level is obtained in the steady state in each individual cell of a population. The stochastic nature of GE gives rise to a distribution of protein levels resulting in a heterogeneous cell population. The heterogeneity termed phenotypic arises even if each cell in the population has identical genetic content, the same initial state and is exposed to the same environment (Kærn et al. 2005).

The subject of stochastic cell biology has grown considerably in the last two decades due to the rapid advancement in single cell experimental techniques, making it possible to link theoretical results with experimental observations (Bressloff 2014). The concepts and methodologies of statistical mechanics, ideally suited for the description of probabilistic phenomena, are used extensively to investigate the effects of stochasticity on the biochemical dynamics of networks (Teschendorff and Feinberg 2021, MacArthur and Lemischka 2013). The cross-disciplinary interaction lays emphasis on quantitative measurements and the elucidation of physical principles at the core of the emergent phenomena. The formalism of statistical mechanics, the foundation of which was laid by Maxwell, Boltzmann and Gibbs in the 19th century, has wide-ranging applicability in the study of natural phenomena. The subject provides an understanding of the macroscopic behavior and the properties of a system (physical, biological, social, economic,…) in terms of the properties of the microstates. A crowning achievement of statistical mechanics in the twentieth century was to develop a comprehensive framework for the study of phase transitions and critical phenomena (Stanley 1999, Yeomans 1992). The concepts and methodologies for the study of phase transitions in equilibrium systems were later extended to the study of nonequilibrium phase transitions. In a thermodynamic system at equilibrium the macroscopic properties do not change as a function of time since there are no net fluxes of matter and energy associated with the system. Nonequilibrium systems are characterized by net flows of matter and energy giving rise to time-dependent physical properties. Living systems are ideal examples of nonequilibrium systems since there are constant exchanges of matter and energy with the outside world (Muñoz 2018). In recent years, a beginning has been made in investigating the possibility of phase transitions and behavior similar to critical phenomena in biological systems. Considerable progress has been made in this direction in the case of neuronal networks describing the dynamics of the human brain (Mora and Bialek 2011, Muñoz 2018). In this review, we focus on some other biological systems which exhibit phase-transition-like behavior in the nonequilibrium. In section 2, we describe noise-induced transitions in the cases of a stochastic model of GE and a model of population genetics. In section 3, we characterize the noise-induced transition exhibited by the population genetics model in terms of critical phenomena. We further describe the critical phenomena associated with the experimentally observed electrochemical signaling in a biofilm which can be understood in terms of the percolation model. Section 4 contains some concluding remarks.

2. Noise-induced transitions

A transition in a system brings about a qualitative change in the state of the system. In the case of deterministic dynamics, such a transition occurs at a bifurcation point (see Box 1). A stochastic system is generally described in terms of a probability distribution of the states. The extrema of the steady state probability density function (PDF), $p_S$, serve as suitable indicators of the qualitative changes associated with a transition (Horsthemke and Lefever 1984). The transition may for example be from a unimodal (PDF has a single peak) to a bimodal (PDF has two peaks) structure of the distribution. The maxima of the PDF represent the most probable values and are analogous to the 'phases' of the system. In a noise-induced transition, as defined above, the deterministic dynamics do not exhibit any qualitative changes across the transition. The transition simply does not occur when there is no noise.

## 2.1 Stochastic Model of GE

We first describe a stochastic model of GE to illustrate a noise-induced transition. In this Random Switch (RS) model, the gene switches between the active (ON) and the inactive (OFF) states at random time intervals. In the ON state the mRNA/protein molecules are both synthesized and degraded whereas only degradation events occur in the OFF state. The dynamics are described by the rate equation

$$\frac{dX}{dt} = j_P z - k_P X \qquad (1)$$

where $X$ represents the protein concentration, $j_P$ is the rate constant for protein synthesis and $k_P$ the protein degradation rate constant. The random variable $z = 1$ (0) in the ON (OFF) state. The only stochasticity is thus associated with the switching of the gene between the ON and OFF states. The synthesis and the degradation of the protein molecules are deterministic events. The variable $X$ in equation (1) may alternatively stand for the mRNA concentration. Define a variable $x = \frac{X}{X_{max}}$ where $X_{max} = \frac{j_P}{k_P}$ is the maximum protein concentration with $0 \leq x \leq 1$. Let $k_a$ and $k_d$ be the rate constants for the transitions between the OFF and ON and the ON and OFF states respectively. One can define two new parameters, $r_1 = \frac{k_a}{k_P}, r_2 = \frac{k_d}{k_P}$, which are the ratios of the protein lifetime, $\tau_P \sim \frac{1}{k_P}$, and the residence times, $\tau_a \sim \frac{1}{k_a}$ and $\tau_d \sim \frac{1}{k_d}$, in the inactive and active states respectively of the gene. In terms of the new parameters the steady state PDF is given by the beta distribution (Karmakar and Bose 2004)

$$p_S(x) = \frac{\Gamma(r_1 + r_2)}{\Gamma(r_1)\Gamma(r_2)} x^{r_1-1} (1-x)^{r_2-1} \qquad (2)$$

where $\Gamma$ represents the gamma function. The beta distribution is a very flexible distribution and can acquire different shapes (i) U-shaped, (ii) bell-shaped, (iii) monotonically increasing, (iv) monotonically decreasing and (v) even straight lines, by changing the shape parameters $r_1$ and $r_2$. The cases (i) - (v) correspond to the parameter values: (i) $r_1 < 1, r_2 < 1$, (ii) $r_1 > 1, r_2 > 1$, (iii) $r_1 > 1, r_2 < 1$, (iv) $r_1 < 1, r_2 > 1$, and (v) $r_1 = r_2 = 1$.

Box 1 Deterministic dynamics

The deterministic dynamics are represented by $N$ coupled differential rate equations where *N* is the number of key biomolecular entities, say, mRNAs and proteins, participating in a biochemical event like gene expression (GE). If there are two genes, the protein products of which regulate each other's expression, *N* would be four with two mRNA and two protein species. A differential rate equation expresses the rate of change in the concentration of a particular biomolecule as the difference in the rates of synthesis and degradation of the biomolecule. Mathematical expressions for the latter two quantities are obtained using appropriate chemical kinetic schemes. The solution of the differential equations, given the initial concentrations, provides knowledge of how the different concentrations evolve as a function of time.

The time evolution is best visualized by drawing trajectories in the state space (Strogatz SH 1994). The state space is $N$-dimensional with each axis representing the concentration values of a specific biomolecule. In the limit of large time, the dynamics may reach a fixed point defining a steady state in which the concentrations attain fixed values, i.e., all the rates of change are zero or fall into limit cycles representing periodic behavior. In the first case, the different trajectories in the state space corresponding to different initial states all converge to the fixed point. In the second case, a trajectory falls into a loop and repeats itself as a function of time. The stability of a fixed point implies that the system regains the fixed point after being weakly perturbed from it. In the case of multisability, there is more than one stable fixed point, each with its own basin of attraction in the state space. The origin of multistability lies in the presence of one or multiple positive feedback loops in the genetic network.

In the case of single-variable dynamics, one can define a potential function $V(x)$ as

$$\frac{dx}{dt} = f(x) = -\frac{dV}{dx}$$

The extrema of the potential function, obtained from $\frac{dV}{dx} = 0$, define the steady states of the dynamics. The minima (maxima) correspond to the stable (unstable) steady states. In the case of bistability, the landscape of the potential function has two valleys (two stable steady states) separated by a hill (the top of the hill represents an unstable steady state). Each valley defines the basin of attraction of the associated stable steady state.

Bifurcations occur in dynamical systems at specific values of the key parameters. In a bifurcation, a change in the dynamical regime occurs in terms of the number of steady states and/or their stability properties. Bifurcations are thus similar to phase transitions which bring about changes in the macroscopic physical properties, e.g., a magnet losing its magnetization above a critical temperature. The most general types of bifurcations are the transcritical, saddle-node, pitchfork and Hopf bifurcations.

Figure 1 shows some of the possible shapes of the beta distribution. The modes (the maxima of the PDF in equation (2)) and the vertical asymptotes ($p'_S(0) = \infty, p'_S(1) = \infty$ with the prime symbol denoting differentiation with respect to $x$) represent the preferential states, the 'phases' (see Box 2) of the system. In the deterministic case the steady states are $x = 0$ and $x = 1$ in the OFF and the ON states respectively. In Case (ii) of the stochastic dynamics, the PDF is unimodal with the mode being given by

$$x_m = \frac{r_1 - 1}{r_1 + r_2 - 2} \qquad (3)$$

The state described by $x_m$ is different from the deterministic stable steady states $x = 0$ and $x = 1$. The appearance of the state $x_m$ is thus noise-induced. In Case (i), the mode $x_m$ becomes the anti-mode (minimum of the PDF). The PDF further has two vertical asymptotes at $x = 0$ and $x = 1$. Though no new states are created, noise gives rise to the coexistence of the two deterministically stable steady states. In Cases (iii) and (iv), the PDF has a single vertical asymptote at $x = 1$ and $x = 0$ respectively (no new states created by noise). In the case of inducible GE activator molecules or transcription factors (TFs) bind the DNA and promote the transition of the gene to the active state. The rate constants $k_a$ and $k_d$ become functions of $S$, the concentration of activator molecules/TFs.

In the parameter regime corresponding to Case (i), one has binary GE in the form of either low ($x = 0$) or high ($x = 1$) protein level. As the concentration $S$ of the activator molecules is increased (decreased), the fraction of cells in the "high" ("low") subpopulation goes up. The response to the changes in the concentration $S$ is thus binary. In the case of Case (ii), on the other hand, a unimodal distribution in the protein levels with graded response is obtained. The maximum of the distribution shifts as the concentration $S$ is changed. The transition from the graded to the binary response is noise-induced (no such transition in the deterministic case) and occurs at the point $r_1 = 1, r_2 = 1$ in the parameter space. At the transition point, the beta distribution reduces to the uniform distribution (flat straight line) which is the maximum entropy distribution. Since entropy is a measure of uncertainty, the state of the system as measured by the random variable $x$ becomes maximally uncertain.

A physical understanding of the unimodal versus binary GE is as follows. In the OFF and ON states of the gene, the stable steady states are $x = 0$ and $x = 1$. If the OFF $\rightarrow$ ON and the ON $\rightarrow$ OFF transition rates, $k_a$ and $k_d$, are much faster than the protein degradation rate ($r_1 > 1, r_2 > 1$), the proteins accumulate over the random transitions resulting in an average protein level intermediate between $x = 0$ and $x = 1$. The probability distribution is unimodal. When the transition rates are much slower than the protein degradation rate

($r_1 < 1, r_2 < 1$), the residence times in the OFF and the ON states are sufficiently long for the protein level to reach the respective steady state values $x = 0$ and $x = 1$.

Box 2. Stochastic dynamics

We consider a single-variable system for which the stochastic rate equation is obtained by adding a noise term to the deterministic rate equation $\frac{dx}{dt} = f(x)$:

$$\frac{dx}{dt} = f(x) + \sigma\, g(x)\xi(t) \qquad (B1)$$

The second term in the above equation represents external noise with $\sigma^2$ being the noise intensity and $\xi(t)$ is a white noise with zero mean and time correlation given by

$$\langle \xi(t)\xi(s) \rangle \sim \delta(t-s) \qquad (B2)$$

Stochastic trajectories in the state space can be drawn by solving the rate equation in equation (B1). Let $p(x,t)dx$ be the time-dependent probability of finding $x$ in the interval $(x, x + dx)$ at time t. An equivalent description of the stochastic rate equation is given by the Fokker-Planck equation (FPE) which directly deals with the probability distributions (Horsthemke and Lefever 1984):

$$\frac{\partial}{\partial t}p(x,t) = -\partial_x[f(x)p(x,t)] + \frac{\sigma^2}{2}\partial_{xx}[g^2(x)p(x,t)] \qquad (B3)$$

The FPE equation can be recast as an equation of continuity expressing the conservation of the probability :

$$\frac{\partial}{\partial t}p(x,t) + \partial_x J(x,t) = 0 \qquad (B4)$$

where $J(x,t)$, interpreted as a probability current, is given by

$$J(x,t) = [f(x)p(x,t)] - \frac{\sigma^2}{2}\partial_x[g^2(x)p(x,t)] \qquad (B5)$$

The stationarity condition, $\frac{\partial}{\partial t}p(x,t) = 0$, results in the further condition $\partial_x J_S(x) = 0$ from equation (B4). The probability current, $J_S(x)$, in the stationary state is thus a constant. The constant is zero in the case of reflecting boundary conditions (zero probability current at the boundaries) resulting in the following equation for the stationary state probability density $p_S(x)$:

$$f(x)p_S(x) = \frac{\sigma^2}{2}\partial_x[g^2(x)p_S(x)] \qquad (B6)$$

Box2. (continued)

> Integrating the above equation, $p_S(x)$ is obtained as
>
> $$p_S(x) = \frac{K}{g^2(x)} exp\left(\frac{2}{\sigma^2}\int \frac{f(u)}{g^2(u)}du\right) \quad (B7)$$
>
> with the integral evaluated at $x$. The pre-factor $K$ is the normalization constant. One can define a stochastic potential, $\varphi(x)$, as
>
> $$p_S(x) = K\, exp\left[-\frac{2}{\sigma^2}\varphi(x)\right] \quad (B8)$$
>
> where
>
> $$\varphi(x) = -\left[\int \frac{f(Z)}{g^2(Z)}dz - \sigma^2 \ln g(x)\right] \quad (B9)$$
>
> with the integral evaluated at $x$. The maxima of $p_S(x)$ correspond to the valleys (minima) of $\varphi(x)$. In the case of additive noise, the function $g(x) = 1$. In this case, the deterministic potential function $V(x)$ (see Box 1) and the stochastic potential $\varphi(x)$ coincide modulo an inessential constant. The extrema of the stationary probability density $p_S(x)$ have an exact correspondence with the valleys and the mountain tops of the potential landscape in the deterministic case. The extrema are thus the stochastic representations of the macroscopic steady states.

This situation corresponds to binary GE. In a cell population binary GE gives rise to phenotypic heterogeneity in the form of two distinct subpopulations. The RS model of GE provides a simple physical understanding of the graded (analog) versus binary (digital) responses in gene regulation (Karmakar and Bose 2004, Munsky and Neuert 2015). In the case of eukaryotic GE, a transition from a closed to an open chromatin structure is required for the transcriptional machinery to access the promoter region of the DNA. The slow rates of these transitions favour the occurrence of binary GE. Raj et al. ((Raj et al. 2006) in their experiment on Chinese hamster ovary cells found evidence that the dominant stochasticity is associated with the random transitions between the OFF and the ON states of the gene. When the variable $X$ in equation (1) represents the mRNA concentration, the steady state PDF described by the beta distribution (equation (2)) reduces to the gamma distribution in the appropriate limits. The distribution arises from bursts of mRNA synthesis during intense ON periods punctuated by OFF periods which have longer durations. The experimental evidence of transcriptional bursts is now quite extensive (Raj and van Oudenaarden 2008, Tunnacliffe and Chubb 2020). The RS model fits the experimental data of Zlokarnik et al. on binary GE quite well (Zlokarnik et al. 1998, Karmakar and Bose 2004). To and Maheshri (To and Maheshri 2010) constructed a synthetic circuit containing a positive feedback loop in budding yeast. The deterministic model of the circuit does not predict bistability. The experimental evidence of bimodal GE induced by noise is in tune with the prediction by the RS model. The RS model has been generalized to include stochastic effects during both transcription and translation and most of the predictions of the simpler model like graded versus binary response and noise-induced transitions continue to remain valid (Shahrezaei and Swain 2008).

## 2.2 Model of Population Genetics

We next discuss another example of a pure noise-induced transition exhibited in a model of population genetics. The deterministic model is described by the rate equation (Horsthemke and Lefever 1984)

$$\frac{dq}{dt} = \frac{1}{2} - q + \lambda q(q-1) \quad (4)$$

where $q$ denotes the state variable with $0 \leq q \leq 1$ and $\lambda$ is the parameter representing the coupling of the system to the environment. The model has a unique stable steady state throughout the parameter regime and displays no bifurcations, i.e., transition from one dynamical regime to another. The parameter λ fluctuates if the system is coupled to a noisy environment with $\lambda = \lambda + \sigma\xi$ where ξ represents a white noise (see Box 2) and $\sigma^2$ is the noise intensity. The associated stochastic differential equation is given by (see Box 2)

$$\frac{dq}{dt} = \frac{1}{2} - q + \lambda q(q-1) + \sigma q(1-q)\xi(t) \quad (5)$$

Comparing with the general form of the stochastic differential equation in Box 2, one can set $f(q) = \frac{1}{2} - q + \lambda q(q-1)$ and $g(q) = q(1-q)$. Following the procedure described in Box 2 the steady state PDF is given by

$$p_S(q) = \frac{K}{[q(1-q)]^2} exp\left[-\frac{2}{\sigma^2}\left(\frac{1}{2q(1-q)} + \lambda ln\left(\frac{1-q}{q}\right)\right)\right] \quad (6)$$

where $K$ is the normalization constant. The extrema, $q_m$, of the PDF satisfy the equation $\left(\frac{dp_S(q)}{dq} = 0\right)$

$$\frac{1}{2} - q_m + \lambda q_m(1-q_m) - \sigma^2 q_m(1-q_m)(1-2q_m) = 0 \quad (7)$$

The roots of the equation for $\lambda = 0$ are given by

$$q_{m0} = \frac{1}{2}, q_{m\pm} = \frac{1}{2}\left[1 \pm \left(1 - \frac{2}{\sigma^2}\right)^{\frac{1}{2}}\right] \quad (8)$$

For $\sigma^2 < 2$, there is only one extremum, a maximum, namely, $q_{m0} = \frac{1}{2}$. When $\sigma^2 > 2$, there are three extrema, a minmum at $q_{m0}$ and two maxima at $q_{m\pm}$. One can thus define a critical noise intensity, $\sigma_C^2$, such that the PDF $p_S(q)$ is unimodal for $\sigma^2 < \sigma_C^2$ and bimodal for $\sigma^2 > \sigma_C^2$. The transition from unimodality to bimodality is purely noise-induced. Figure 2 shows the plots of $p_S(q)$ versus $q$ (equation (6)) for $\sigma^2 < \sigma_C^2, \sigma^2 = \sigma_C^2$ and $\sigma^2 > \sigma_C^2$. At the critical noise intensity the probability distribution is flat-topped. The stochastic potential (see equation $(B9)$, Box 2) is given by

$$\varphi(q) = \frac{1}{2q(1-q)} + \sigma^2 ln[q(1-q)] \quad (9)$$

Figure 3 shows the plot of the stochastic potential versus $q$ for $\sigma^2 > \sigma_C^2$. In this case the potential has two minima. The potential has a single minimum when $\sigma^2 < \sigma_C^2$.

### 3. Critical-like Phenomena

In the vicinity of the noise-induced critical point, power-law behavior emerges in analogy with similar behavior in the cases of equilibrium and nonequilibrium critical point transitions (see Box 3). Some analogies can be drawn between bifurcations and equilibrium thermodynamic phase transitions. The change in the dynamical regime at a bifurcation point is similar to a phase transition. Bifurcations in dynamic systems are governed by deterministic dynamics unlike in the cases of equilibrium thermodynamic phase transitions which are driven by thermal fluctuations. In the mean-field limit of the latter type of transitions, thermal fluctuations are ignored through the replacement of a fluctuating quantity by its average value. Once this limit is reached, the critical exponents (see Box 3) become independent of the dimension of the system. A mean-field critical system is similar to a dynamical system near its bifurcation point as will be shown in the following.

Box 3. Phase transitions and critical phenomena

> Well-known examples of equilibrium phase transitions are the liquid-gas, paramagnet-ferromagnet (PM-FM) and superconductor-metal phase transitions. In such transitions one can identify an order parameter which has a non-zero (zero) expectation value below (above) a critical temperature $T_C$. In the case of the liquid-gas and PM-FM transitions, the order parameters are the density difference in the two phases and the spontaneous magnetization respectively. In a first-order transition, the order parameter jumps discontinuously at the transition point. In a second-order (continuous) transition, the order parameter decreases continuously to a zero value at the transition point, termed the critical point. In the cases of the liquid-gas and PM-FM transitions, a line of first-order transitions, in the pressure-temperature and magnetic field-temperature phase diagrams, respectively, terminates at a critical point. The line separates two distinct phases, liquid and gas, magnetic and non-magnetic, in the phase diagram with the distinction between the phases disappearing at and beyond the critical point (Yeomans 1992).

Box 3. (continued)

The critical point transition is characterized by emergent universal features which are collectively described as critical phenomena. In the vicinity of a critical point transition thermodynamic quantities, say the order parameter, acquire a general power-law form ~ $|(T - T_C)|^\theta$ where $T$ and $T_C$ represent the absolute temperature and the critical temperature respectively. The exponent $\theta$ is the critical exponent which has different values, positive or negative, for different thermodynamic quantities. Critical point transitions, occurring in two different systems, belong to the same universality class if the number of components of the order parameter and the dimension of the embedding space are the same. In this case the numerical values of the exponents associated with equivalent thermodynamic quantities are identical. The liquid-gas transition and the PM-FM transition, as described by the three-dimensional Ising model, belong to the same universality class. Universality implies that close to the critical point microscopic details of the systems are irrelevant. Long-range spatial correlations emerge in the system, close to the critical point, with power-law divergence of the correlation length. The critical phenomena constitute one of the finest examples of emergence in a physical system. The formalism of equilibrium phase transitions has been extended over the years to include diverse types of phase transitions like nonequilibrium phase transitions, self-organised criticality (SOC) ) and the percolation transitions (Sornette 2006, Stauffer and Aharony 2003). In the case of SOC, no external tuning of the system (say, by varying the temperature) to the vicinity of the critical point is needed to observe the critical behavior. The inherent dynamics of the system bring it to the self-organised critical state distinguished by the appearance of power-laws.

The percolation model describes phase transitions in disordered systems. Consider a regular network, e.g., a square lattice. A disordered network is generated by requiring that the probability that a site (site percolation) or a bond (bond percolation) is occupied is $p$. If $p = 1$, there is full connectivity. If $p = 0$, the network disappears. A percolation transition occurs at a critical value, $p_C$, establishing long-range connectivity across the system. In the limit of infinite size, each network has a unique value of $p_C$ and a sharp transition is obtained. The long-range connectivity is described in terms of an infinite cluster of

Box 3. (continued)

> connected sites or bonds. The cluster is absent (present) for $p < p_C$ ($p > p_C$) suggesting two distinct phases. Close to the percolation threshold probability $p_C$, cluster-related quantities acquire power-law forms $\sim |p - p_C|^\theta$. The critical phenomena are similar to those in the case of equilibrium thermodynamic phase transitions. In the case of the percolation transition, the role of temperature is played by the occupation probability $p$.

The presence of a positive feedback in a biochemical system is known to promote bistability with the transition from monostability to bistability occurring at a bifurcation point. In the case of the saddle-node (SN) bifurcation (Strogatz 1994), the transition is like a first-order transition with a discontinuous change in the magnitude of the dynamical variable. At the bifurcation point of a supercritical pitchfork (SP) bifurcation, the transition occurs without discontinuities at the bifurcation point and is thus analogous to a second-order or continuous phase transition (Strogatz 1994). We designate this transition as a critical point transition. Figure 4 illustrates the two types of bifurcation: saddle-node and pitchfork. Recently, Erez et al. (Erez 2019) showed that biochemical models with positive feedback, of relevance in several cell biological processes, exhibit power-law behavior close to a critical-point transition. The critical behavior belongs to the universality class of the mean-field Ising model. The Ising model is the simplest statistical mechanical model which illustrates a magnetic transition from the paramagnetic to the ferromagnetic state. The mapping between the equilibrium and nonequilibrium models close to their critical points enables one to express the thermodynamic quantities like the order parameter (see Box 3), temperature, magnetic field, susceptibility and heat capacity in terms of the biochemical parameters. Erez et al. (Erez et al. 2019) verified their quantitative theoretical predictions in single cell experiments on T-cell signaling.

Bose and Ghosh (Bose and Ghosh 2019), have demonstrated the equivalence between the biochemical models with positive feedback and the Ising mean-field model, in the vicinity of their respective critical points, using a simple deterministic approach. As illustrations of the method a model of population dynamics incorporating the strong Allee effect and the two-variable genetic toggle were considered. The method works when there is no noise-induced transition in the full stochastic model, i.e., the extrema of the PDF in the stochastic case are in one-to-one correspondence with the deterministic steady states. This holds true in the case additive noise, $g(x) = 1$ (see Box 2), and for sufficiently small multiplicative noise, $g(x) \neq 1$. We now show that the same method, as proposed earlier, can be used in the case of a noise-induced transition with the starting equation being different (Bose 2022). In the deterministic case, one starts with the steady state equation, $F(x^*) = 0$, obtained from the rate equation, $\frac{dx}{dt} = F(x)$, with $x^*$ representing the steady states. In the case of a noise-induced transition, the starting equation is the one satisfied by the extrema of the steady state PDF of the stochastic model. We consider the case of the population genetics model for which a noise-induced transition occurs, as discussed in section 2.2. From equation (7) we obtain an equation analogous to the deterministic steady state equation as

$$F(q_m) = \frac{1}{2} - q_m + \lambda q_m(1 - q_m) - \sigma^2 q_m(1 - q_m)(1 - 2q_m) = 0 \quad (10)$$

The normal form of a dynamical rate equation exhibiting a SP bifurcation is given by (Strogatz 1994, Guckenheimer and Holmes 2013)

$$\frac{dq}{dt} = rq - q^3 \quad (11)$$

where $r$ is the bifurcation parameter. The rate equation is symmetric under the transformation $q \to -q$. The symmetry is lost in the case of the imperfect SP bifurcation for which the rate equation includes an extra term $H$ in the right hand side of equation (11). The steady state equation in this case is (Strogatz 1994)

$$rq - q^3 + H = 0 \quad (12)$$

Equation (12), valid in the vicinity of the bifurcation point, is identical in form to the mean-field equation of state of the Ising model close to its critical point (Yeomans 1992, Goldenfeld 1992), namely,

$$h - \theta m - \frac{1}{3}m^3 = 0 \quad (13)$$

where $m$ represents the average magnetization per spin, $\theta = \frac{(T-T_C)}{T_C}$ is the reduced temperature with $T_C$ being the critical temperature and $h$ is a dimensionless magnetic field. The critical point of the Ising model is the point, $\theta = 0, h = 0$, in parameter space. The order parameter, spontaneous magnetization $m$ (with $h = 0$), becomes zero at the critical point (see Box 3). To obtain correspondence with the Ising model, we reduce the expression $F(q_m) = 0$ (equation (10)) to the normal form, equation (12), of the imperfect SP bifurcation. The procedure is standard (Bose and Ghosh 2019): Taylor expand the function $F(q_m)$ around the point $q_C$ to third order with the point $q_C$ fixed by the condition $F''(q_C) = 0$ so that a second-order term does not appear in the expansion consistent with the normal forms in equations (12) and (13). The Taylor expansion is given by

$$F(q_C) + F'(q_C)(q_m - q_C) + F'''(q_C)\frac{(q_m - q_C)^3}{3!} = 0 \quad (14)$$

The condition can be put in the normal form, as in equation (12), with

$$q = \frac{(q_m - q_C)}{q_C}, H = -\frac{6F(q_C)}{F'''(q_C)q_C^3}, r = -\frac{6F'(q_C)}{F'''(q_C)q_C^2} \quad (15)$$

Comparing equations (12) and (13), one finally obtains,

$$m = \frac{(q_m - q_c)}{q_c}, h = -\frac{2F(q_c)}{F'''(q_c)q_c^3}, \theta = \frac{2F'(q_c)}{F'''(q_c)q_c^2} \quad (16)$$

At the critical point, the magnetization $m = 0$, i.e., $q_c = q_m$. In the case of the population genetics model $q_c$ is given by

$$q_c = \frac{1}{2} - \frac{\lambda}{6\sigma^2} \quad (17)$$

The critical point is given by $\lambda = 0$ so that equation (10) has a triple root given by $q_m = q_{m0} = q_{m\pm} = q_c = \frac{1}{2}$ (see equation (8)), with the noise intensity $\sigma^2 = 2$ at the critical point. Analysing equation (13), one can obtain power-law forms for the various quantities associated with the Ising model (see Box 3). With the field $h = 0$, the order parameter $m = 0$ when the reduced temperature $\theta$ is positive and $m \sim \pm(-3\theta)^\beta$ for $\theta$ negative. The critical exponent has the mean-field value $\beta = \frac{1}{2}$. For $\theta = 0$, $m \sim (3h)^{1/\delta}$ with the critical exponent $\delta = 3$. The critical exponents associated with the thermodynamic quantity susceptibility, $\chi \equiv (\partial_h m)_{h=0}$, are $\gamma = 1$ for both $\theta > 0$ and $\theta < 0$. In the case of the population genetics model, the quantities $q, H, r$ are analogous to the quantities $m, h, \theta$ in the magnetic model. The numerator in the expression for $r$ (equation (15)) involves the quantity $F'(q_c)$. One can check that at the critical point ($\lambda = 0$, $\sigma^2 = 2$) the conditions $F(q_c) = 0$, $F'(q_c) = 0$ are obeyed so that the quantities $H$ and $r$ in equation (15) are zero as they should be in analogy with the magnetic model. Also, $F'(q_c)$ changes sign at the critical noise intensity $\sigma^2 = 2$, having a positive (negative) value for noise intensity greater (lesser) than the critical noise intensity. The quantity $F'''(q_c)$ in the denominator of $r$ is always negative. The quantity $r$ thus changes sign at the critical noise intensity just as the parameter $\theta$ in the magnetic model changes sign at the critical temperature $T_C$. The noise intensity in the population genetics model is analogous to temperature in the magnetic model. In analogy with the magnetic model, the critical behavior in the population genetics model is given by the relations, with the parameter $\lambda$ playing the role of the magnetic field,

$$q \sim (\sigma^2 - \sigma_C^2)^{\frac{1}{2}}, \lambda = 0, \sigma^2 > \sigma_C^2 \quad (18)$$
$$q \sim \lambda^{1/3}, \sigma^2 = \sigma_C^2 = 2 \quad (19)$$
$$\left(\frac{\partial q}{\partial \lambda}\right)_{\lambda=0} \sim |\sigma^2 - \sigma_C^2|^{-1} \quad (20)$$

The critical exponents are identical to those obtained in the case of the magnetic model so that the two different models belong to the same universality class. This further shows that a mean-field equilibrium thermodynamic phase transition and a nonequilibrium noise-induced transition share a close kinship in the vicinity of the critical point. As discussed in Box 3, some equilibrium phase transitions are characterized by a line of first-order phase transitions terminating at a critical point. The $H$ versus $r$ phase diagram corresponding to equation (12) shows a similar feature. Two lines of SN bifurcations or first order transitions, enclosing a region of bistability, merge at the critical point $r = 0, H = 0$ (Strogatz 1994, Bose and Ghosh 2019).

We next describe another critical point transition in a biological system which is well-described by the percolation model (see Box 3) of statistical physics. A recent experiment provides evidence of long-range electrochemical signaling in a *Bacillus Subtilis* biofilm (bacterial community), brought about by nutrient starvation (Prindle et al. 2015). The molecular mechanism governing the signal propagation is as follows. During the expansion of a biofilm the nutrient glutamate required for growth is mostly utilized by the peripheral cells so that the glutamate does not reach the interior. This results in the opening up of the potassium channel of a stressed interior cell through which potassium ions are released outside. The local increase in the concentration of the extracellular potassium ions depolarizes the neighbouring cells preventing their uptake of the charged amino acid glutamate. A depolarized cell, experiencing nutrient stress, opens up its own potassium channel for the release of the potassium ions. A succession of depolarizing events, occurring as a chain reaction, propagates the electrochemical signal to the biofilm periphery. At the edge, the signal stops the growth of the outermost cells by imposing a limitation on the glutamate uptake. The glutamate is then able to reach the starved bacterial cells in the interior of the biofilm thereby reducing the nutrient stress. The signaling ceases at this point so that the bactetial biofilm restarts its growth. Due to the renewed growth process the interior cells are once more deprived of nutrients. The cycle of events (starvation of interior cells, initiation of electrochemical signaling, halting of biofilm expansion etc.) repeats itself giving rise to oscillations in the biofilm growth rate. The signaling process offers protection to the biofilm against external attacks by maintaining a population of viable cells at the core thus conferring a population-level benefit. The community-level benefit is, however, accompanied by a cost to the individual cells which participate in the firing (signaling) process. The cost is in terms of reduced growth rates due to the metabolic burden of the signaling activity.

The question that arises naturally is how the biofilm resolves the cost-benefit problem. The answer is provided by the percolation model, a model of phase transitions in spatially heterogeneous systems (see Box 3) (Stauffer and Aharony 2003). The heterogeneity in the biofilm arises from the fact that only a fraction of the cells participates in the signaling process. A recent study has shown (Larkin et al. 2018) that the cost-benefit tradeoff organizes the ensemble of firing cells to operate at the critical percolation point. Let $p$ be the probability that a cell is a firing cell. The probability is a measure of the fraction of firing cells in the ensemble of cells. Long range electrical signaling across a biofilm is possible only if a connected path of firing cells spans the biofilm. A spanning path exists only if the condition $p \geq p_C$, the percolation critical probability, is satisfied (see Box 3). The collective benefit and the collective cost of the signaling can be quantified using appropriate measures. The ratio benefit/cost was found to reach its peak value at the percolation threshold $p_C$ (critical point). The experimentally measured fraction of firing cells agrees with the theoretically predicted value of the percolation threshold for a triangular lattice (each bacterial cell in the biofilm has roughly six nearest neighbours ). The firing cells constitute clusters of various sizes. In a cluster the firing cells are located at nearest-neighbour positions. Percolation theory predicts that close to the percolation threshold, the cluster size distribution has a power law form, $D(s) \sim s^{-\tau}$ where $\tau$ defines the critical exponent. The experimentally measured value of $\tau$ is approximately equal to 2, the theoretically predicted value in the percolation model. There are suggestions that biological networks, like the neuronal and gene regulatory networks, operate at criticality separating the ordered (characterized by robustness to perturbations) and the disordered phases (flexible responses) (Mora and Bialek 2011). The experimental observations on a biofilm, subjected to nutrient stress, demonstrate the advantage of the system to be poised at a critical point since at this point the benefit outweighs the cost of long-range electrochemical signaling.

4. Discussion

In this review, we have mainly discussed noise-induced transitions and critical phenomena in biological systems using statistical mechanical approaches. The transitions considered in a stochastic GE model and a population genetics model are "purely" noise-induced, i.e., in the absence of noise (deterministic limit), there are no transitions. The utility of the RS model of gene expression is that it is simple enough to be analytically tractable but at the same time provides a clear understanding of some significant features of stochastic GE, namely, binary gene expression, analog versus digital response to changing inducer concentrations and a noise-induced bimodal GE. Each of the phenomena described has considerable experimental support. The noise intensity $\sigma^2$ plays the role of a bifurcation parameter, in the case of the population genetics model, driving a transition from a unimodal to a bimodal steady state PDF at a critical value of the noise-intensity. The description "purely" is emphasized as noise-driven transitions among deterministically pre-existing steady states are well-known. In the presence of positive feedback loops the deterministic dynamics exhibit multistability in specific parameter regimes. In the case of bistability the potential landscape exhibits two minima (the valleys representing two stable steady states) separated by a maximum (the hilltop representing an unstable steady state) (see Box 1). When the noise intensity is comparable to the barrier height separating the two minima, the system switches from one valley state to the other. Statistical mechanics formalisms, covering both analytical and numerical approaches, enable one to determine the steady state PDF as well as quantities like the mean first passage time from one potential well to the other. A large number of single-cell experiments combined with statistical mechanical concepts and techniques elucidate the role of noise-driven phenomena in biological systems.

Most of the noise-driven phenomena are not poised at criticality. The cell exploits noise to generate phenotypic heterogeneity, advantageous for the cell population to survive under stress. This is the well-known bet-hedging strategy adopted by microorganisms like bacteria and viruses to cope with the stresses like nutrient depletion, exposure to antibiotic drugs etc. (Balaban NQ 2011). In general, stochastic bistable dynamics (deterministically bistable plus noise-driven transitions between the two wells of the potential landscape), promote the coexistence of two phenotypically distinct subpopulations of cells. The level of a key regulatory protein is low (high) in subpopulation 1 (subpopulation 2). The subpopulation 1 of a bacterial population dies under stress whereas in subpopulation 2, due to the high level of the key regulatory protein, a chain of biochemical events is initiated resulting in, say, the reduction of the bacterial growth rate to a very low value. Antibiotics usually target growing cell walls. Also, nutrients are required for bacterial growth. Subpopulation 2 with a negligible growth rate is thus able to survive under nutrient depletion or antibiotic treatment. The "normal" bacteria, in Subpopulation 1, grow at normal rates and are not able to survive under stress.  It is the existence of the "persisters", constituting subpopulation 2, which prevents the full eradication of a microbial infection. The persisters wait for an opportune time (the treatment is over) to switch to the normal mode of growth and the infection is reinitiated. Experimental evidence of a persister subpopulation has been found in a number of species from *E.coli* to mycobacteria (Bálazsi et al. 2011, Sureka et al. 2008, Ghosh et al. 2011). Stochastic fluctuations play an important role in cell-fate decisions. There is now experimental evidence that certain chemical compounds and biochemical complexes can function as noise modulators (noise enhancers) without changing the mean expression level (Dar et al.2014, Desai et al. 2021). By regulating the noise intensity, it is possible to shift the decision-making process towards a specific fate, e.g., reactivating  HIV from latency (a quiescent state  which prevents HIV cure) to the active-replication state in which the virus can be eliminated through appropriate therapy.

By changing the noise intensity through the use of chemical molecules, the detection of noise-induced transitions would be possible in appropriately-designed experiments.

In this review, we have used the description "critical" to describe continuous transitions, as occurs, for example at the SP bifurcation point (figure 4). The regime shifts which occur at a SN bifurcation point are termed discontinuous. A large body of literature exists on the early signatures of sudden regime shifts in dynamical systems (ecological, environmental, financial, biological…). The sudden changes occur at the SN bifurcation points, termed in the literature, as critical point or tipping point transitions. A few examples of the early signatures of sudden regime changes are the critical slowing down, a rising variance and an increasing autocorrelation function which reaches its maximum value at the "tipping" point. The early signatures of regime shifts at the bifurcation points involving continuous changes are similar. We refer the reader to some reviews on the rapidly growing subject (Scheffer M et al. 2009, Scheffer M et al. 2012).

There are some recent suggestions that cell differentiation is akin to a phase transition from the pluripotent state to the differentiated states representing distinct cellular types. The differentiation process has a binary character with a progenitor cell differentiating into either of two possible lineages. For example, a blood cell differentiates into the erythroid and myeloid lineages. The early signatures of the "transition" at a SP bifurcation point separating the undifferentiated and differentiated cellular phases have been computed by Pal et al. (Pal et al. 2015). Mojtahedi et al. (Mojtahedi 2016) have extended the earlier results and provided experimental verification of some of the theoretical proposals. Statistical mechanical studies of the differentiation process provide a framework for the further exploration of the idea of cell differentiation as a phase transition (Garcia-Ojalvo and Arias 2012, Ridden et al. 2015, Bose and Pal 2017).

One of the first proposals that a functional biological network may be tuned at a critical point was made in the case of a gap-gene network in a developing fruit fly embryo (Krotov et al. 2014). A careful analysis of the gene expression data provides some characteristic signatures of criticality with the underlying dynamics involving a pair of mutually repressive gap genes. A similar two-gene motif is at the core of the network governing cell differentiation and one obtains similar critical point signatures close to the SP bifurcation point at which cell differentiation occurs in the theoretical model (Bose I and Pal M 2017). Another biological process which provides experimental signatures of criticality is that of tissue homeostasis (Clayton et al. 2007, Rué and Arias 2015). In the adult tissue, the homeostatic condition is achieved when the number of progenitor ($P$) cells capable of proliferation as well as differentiation remains constant so that the tissue size is unchanging. The $P$ cells undergo cell division with three possible outcomes: $PP$ (both the daughter cells are $P$ cells), $PD$ (one daughter is a $P$ cell while the other is a differentiated ($D$) tissue cell) and $DD$ (both daughter cells are differentiated cells), the probabilities of the outcomes being $a, b$ and $c$ respectively with $a + b + c = 1$. The condition for homeostasis is $a = c$ which is satisfied when the proliferation rate of the $P$ cells is counterbalanced by the loss rate due to differentiation. The colony of $P$ cells which grows from a single progenitor cell defines a branching process with three distinct dynamical regimes: subcritical ($a < c$), critical ($a = c$) and supercritical ($a > c$). The theory of branching processes, in nonequilibrium statistical mechanics, is well-developed with the availability of several powerful theorems and rigorous results (Harris 1963, Kimmel and Axelrod 2015). The applications of the theory range from cosmic ray showers, nuclear chain reactions to epidemic spread and the growth/extinction of reproducing populations including tumours (Remy and Cluzel 2016). The branching process models have extensive applications in biology including oncology (Kimmel and Axelrod 2015, Durrett 2015). There is now considerable theoretical and experimental support to the idea that a critical branching process provides a good description of tissue homeostasis in the adult mammalian skin (Clayton et al. 2007, Rué and Arias 2015, Ghosh and Bose 2020). The analogies between a critical branching process and the mean-field percolation and sandpile ( a sandpile model is a toy model for self-organised criticality (SOC) mentioned in Box 3) models, close to criticality, are well-known ( Alstrøm 1988, Zapperi et al. 1995).

The noise-induced transitions and critical point phase transitions discussed in this review are examples of emergent phenomena in biological systems. In section 2, the noise-induced transitions discussed occur in zero-dimensional systems. Since the systems are not spatially extended, the transitions lack some of the generic features of equilibrium phase transitions, hence the omission of the term "phase" in the description "noise-induced transitions". Noise-induced nonequilibrium phase transitions, pertaining to spatially extended systems, are known to exhibit the full range of critical phenomena including the divergence of the correlation length and the breaking of symmetry and ergodicity (Van den Broeck et al. 1994, Toral 2011). The inclusion of living matter within the ambit of natural systems, exhibiting a variety of transitions, highlights the universal features of emergent phenomena across the living-nonliving divide. The excitement and challenge lie in the design of novel experiments which capture the essential ideas embodied in the statistical mechanical descriptions. A beginning in this direction has already been made (Muñoz 2018, Teschendorff and Feinberg 2021) and one expects to witness a surge in research activity in the coming years, offering new insights and novel perspectives.

Acknowledgement. IB acknowledges the support of NASI, Allahabad, India under the Honorary Scientist Scheme. IB also thanks Achintya Singha for his help in preparing the manuscript.

Glossary: Basic Terminologies

Dynamical system: a system the state of which, defined in terms of the magnitudes of some key variables, changes as a function of time.

State space: consists of all possible states. In the case of an $N$-variable system, the state space is $N$-dimensional with one coordinate axis for each variable.

Bistability: coexistence of two stable steady states for the same parameter values.

Bifurcation: a change in the dynamical regime occurring at specific parameter values. The change is in terms of the number and/or the stability properties of the steady states obtained as solutions of the differential rate equations.

Equilibrium system: the macroscopic properties of the system do not change as a function of time. There are no net fluxes of matter and energy associated with the system.

Nonequilibrium system: characterized by net flows of matter and energy resulting in time-dependent physical properties.

Stochastic process: involves an element of uncertainty. The process is described in terms of random variables the time evolution of which is due to chance fluctuations (noise).

Ergodicity: time average of a measurable quantity is equal to the average taken over the state space of the system.

Allee effect: in the case of the strong Allee effect, the per capita growth rate of a population is negative below a threshold population size or density resulting in an extinction of the population. Above the threshold, the per capita growth rate is positive and reaches its maximum value at an intermediate density.

Additive/multiplicative noise: additive noise does not depend on the state of the system, multiplicative noise does.

Noise-induced transition: no transition occurs in the case of deterministic dynamics. Multiplicative noise brings about a transition from one dynamical regime to another. An example of regime change is the change from a unimodal steady state probability distribution to a bimodal one at specific parameter values.

Phase transition: may occur in the equilibrium as well as in the nonequilibrium. Sudden qualitative changes occur in the physical properties of a system at the transition point defined by specific parameter values.

Order parameter: a physical quantity which has a non-zero expectation value in one phase and zero expectation value in the other phase.

Critical point transition: the order parameter, in the ordered phase of the system, goes continuously to zero at the combination of parameter values defining the critical point.

Critical phenomena: phenomena occurring in the vicinity of the critical point. One of these is the development of power-law singularities in various quantities like the order parameter, susceptibility, specific heat etc. as the critical point is approached.

Self-organised criticality: the inherent dynamics of the system lead it to the critical state, no external tuning to the critical point is required.

Percolation model: a model which addresses the issue of long-range connectivity in a disordered system. In the case of a disordered lattice, the disorder is in terms of missing sites or bonds, with probability $p$, at random locations. The critical point is defined by the critical percolation probability, $p_C$, above which long-range connectivity exists across the system. Critical phenomena are exhibited by cluster-related quantities as the critical point is approached.

Figure Legends

Figure 1. The beta distribution describing the steady state PDF of the RS model of stochastic GE (equation (2)). Different shapes of the distribution are obtained in different parameter regimes.

Figure 2. The plots of the steady state PDF, $p_S(q)$ versus $q$ (equation (6)), in the population genetics model for different values of the noise intensity $\sigma^2$. The distribution is unimodal for $\sigma^2 < \sigma_C^2 = 2$, flat-topped for the critical noise intensity $\sigma^2 = \sigma_C^2$ and bimodal for $\sigma^2 > \sigma_C^2$.

Figure 3. The stochastic potential, $\varphi(q)$ versus $q$ (equation (9)), in the population genetic model for the noise intensity $\sigma^2 > \sigma_C^2$, the critical noise-intensity. The potential has two minima reflecting the bimodal character of the PDF.

Figure 4. The saddle-node (SN) (left figure) and the supercritical pitchfork (SP) (right figure) bifurcations. In the left figure, a pair of SN bifurcations separates a region of bistability (shades grey) from two regions of monostability. The solid branches represent stable steady states and the dashed branch an unstable steady state. The steady state changes discontinuously at the bifurcation points $u_1$ and $u_2$ (analogous to a first-order transition) and exhibits hysteresis. In the case of the SP bifurcation, the solid branches represent stable steady states and the dotted branch an unstable steady state. The bifurcation point separates a monostable regime from a bistable one, with changes occurring continuously at the bifurcation point (analogous to a second-order or critical point transition).

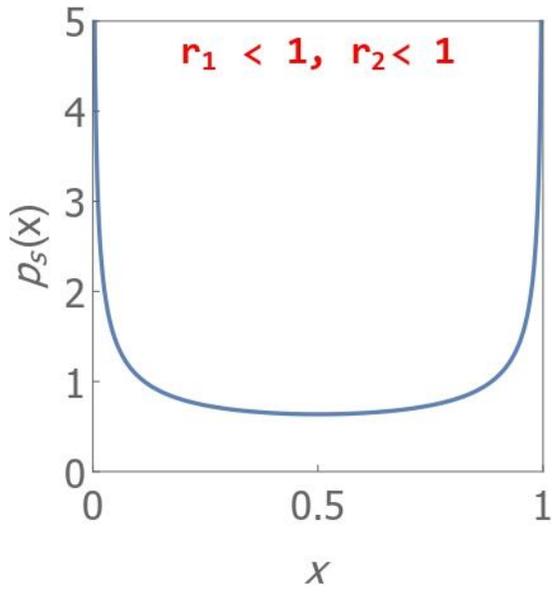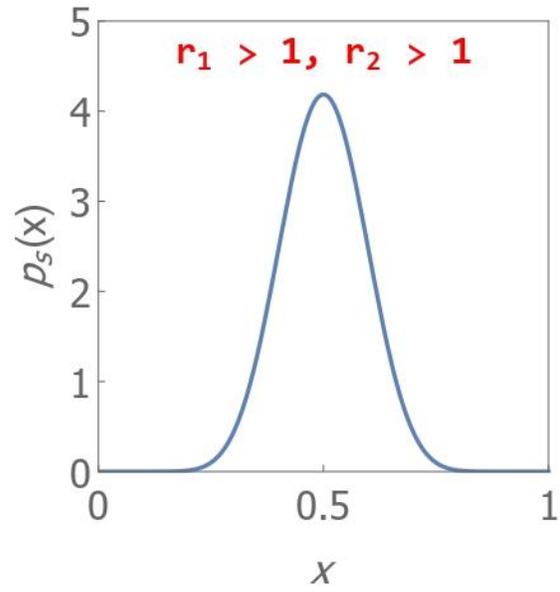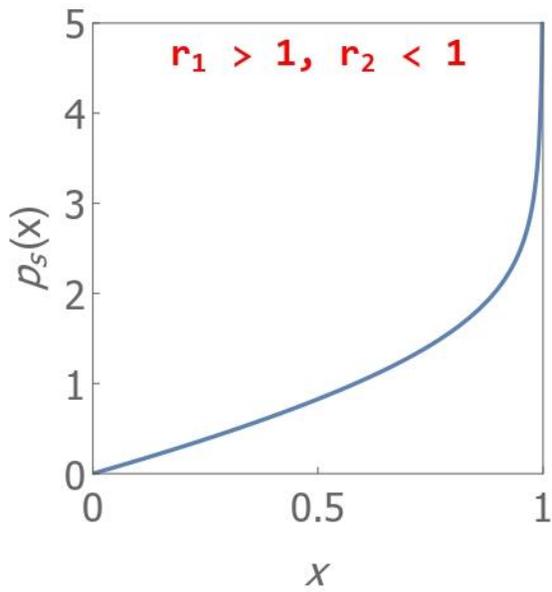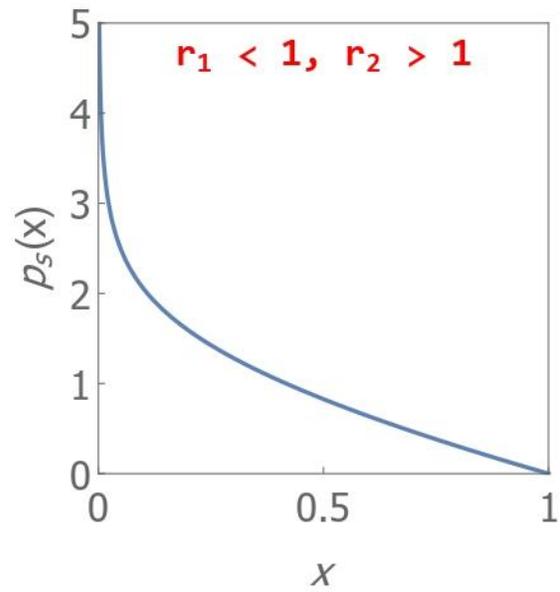

Figure 1

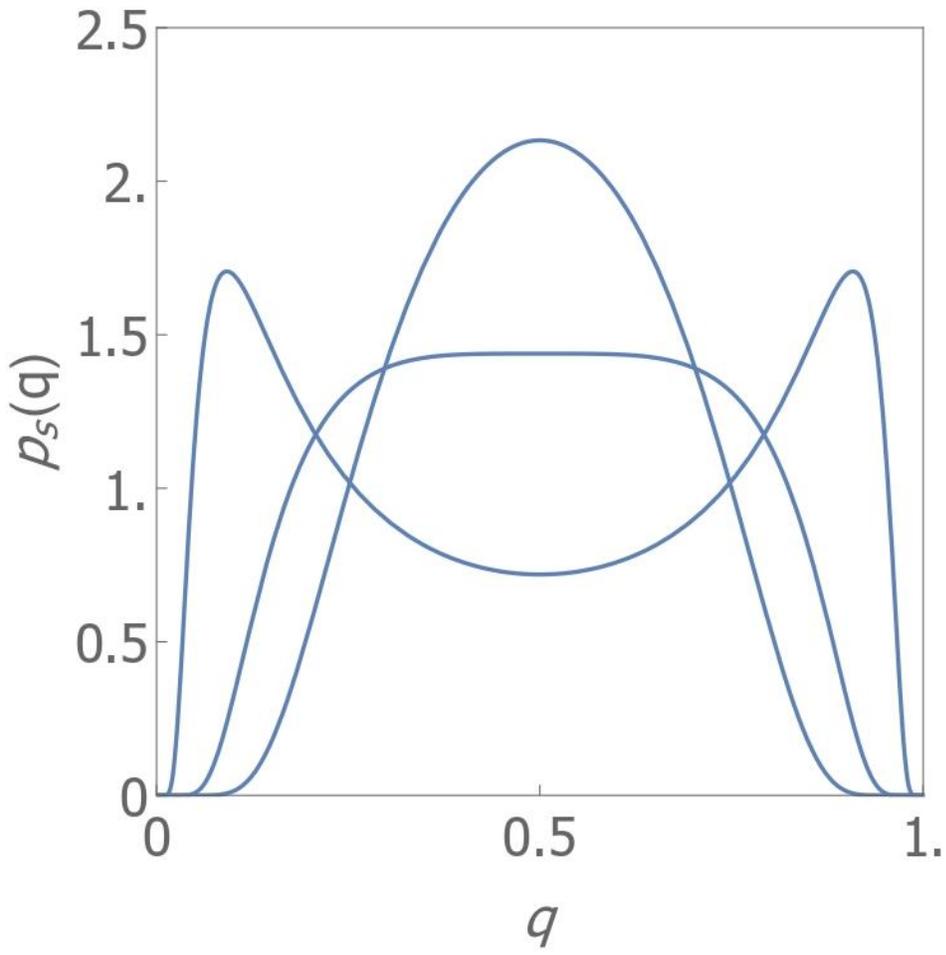

Figure 2

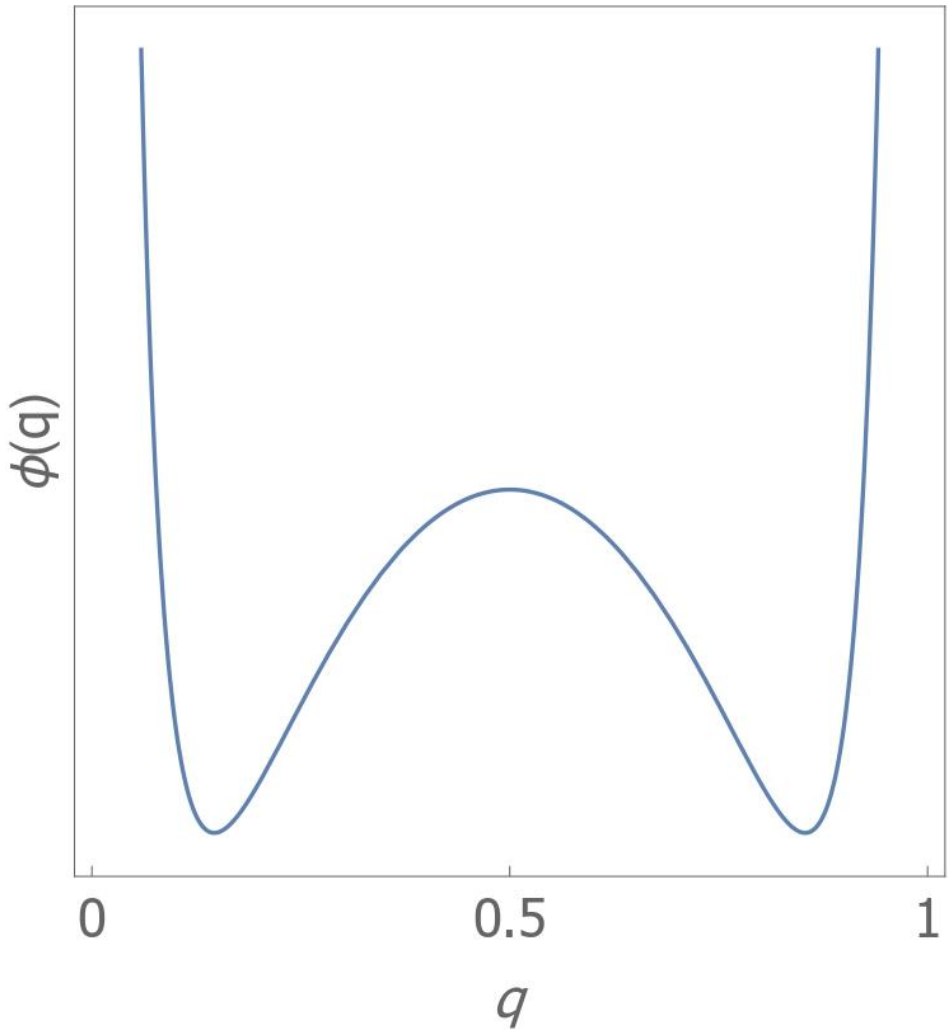

Figure 3

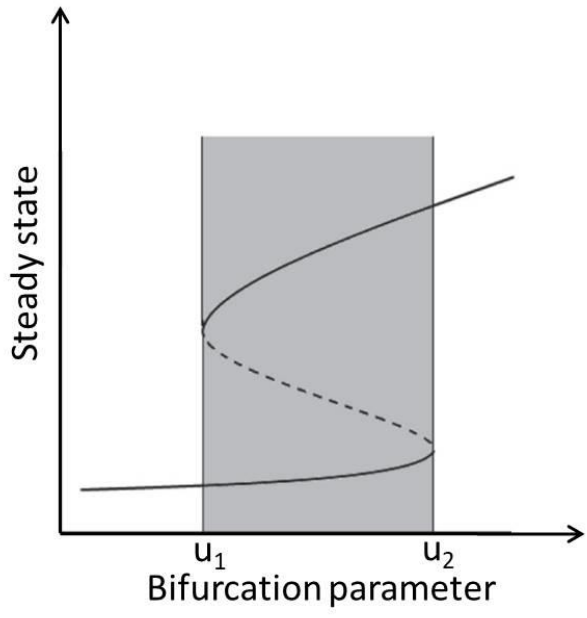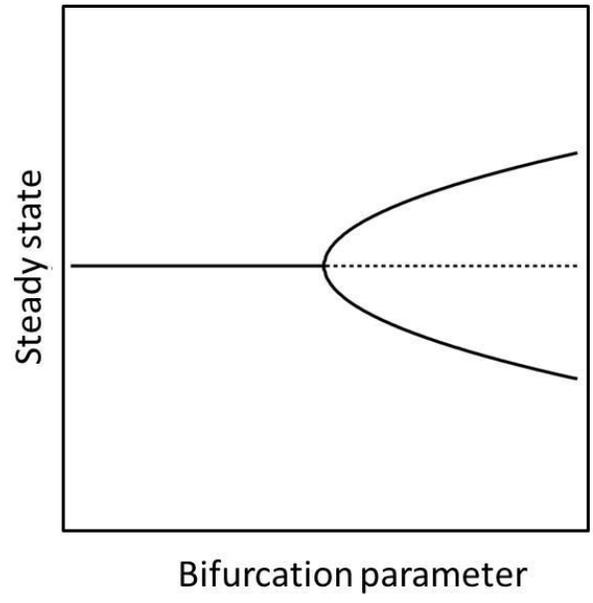

Figure 4